\documentclass[aps,pra,preprint,a4paper,superscriptaddress,floatfix]{revtex4-2}

\usepackage[left=13.5mm,right=13.5mm,top=17mm,bottom=34mm]{geometry}

\usepackage{times}
\usepackage{graphicx,xcolor}
\usepackage{amsmath,amsfonts,amssymb}
\usepackage{bm}
\usepackage{listings}
\usepackage{soul}

\usepackage{booktabs}
\usepackage{multirow}

\usepackage{afterpage}

\usepackage[
unicode=true,
bookmarks=true,%
bookmarksnumbered=true,%
hyperfootnotes=false,
colorlinks=true,%
linkcolor=blue,%
citecolor=blue,%
urlcolor=blue
]{hyperref}  

\newenvironment{mat}{\left[\begin{array}{ccccccccccccccc}}{\end{array}\right]}
\newcommand\bcm{\begin{mat}}
\newcommand\ecm{\end{mat}}

\newenvironment{cmat}{\left(\begin{array}{ccccccccccccccc}}{\end{array}\right)}
\newcommand\bcrm{\begin{cmat}}
\newcommand\ecrm{\end{cmat}}

\newenvironment{rmat}{\left[\begin{array}{rrrrrrrrrrrrr}}{\end{array}\right]}
\newcommand\brm{\begin{rmat}}
\newcommand\erm{\end{rmat}}

\definecolor{red1}{HTML}{d21e1e}
\definecolor{blue2}{HTML}{0D5EAF}
\definecolor{blue1}{RGB}{90, 200, 255}
\definecolor{pink1}{RGB}{255, 180, 180}
\definecolor{green1}{RGB}{0, 125, 0}
\definecolor{green2}{RGB}{150, 220, 150}
\definecolor{orange1}{HTML}{faaa82}
\definecolor{purple1}{HTML}{d29bf5}
\definecolor{purple2}{HTML}{9c00ff}

\usepackage{titlesec}
\titleformat{\paragraph}[runin]
  {\normalfont\normalsize\itshape}{}{}{}[\,---\,]
\titlespacing*{\paragraph}{0pt}{*0}{0pt}
\titleformat{\subparagraph}[runin]
  {\normalfont\normalsize}{}{}{}[:]

\newcommand{\mstitle}
{Formation of mechanical rogue waves}

\begin{document}

\title{\mstitle}

\author{Yasuhiro Miyazawa}
\affiliation{Department of Mechanical Engineering, Seoul National University, 1 Gwanak-ro, Gwanak-gu, Seoul 08826, South Korea}
\affiliation{Department of Aeronautics and Astronautics, University of Washington, Seattle, Washington 98195}
\author{Christopher Chong}
\affiliation{Department of Mathematics, Bowdoin College, Brunswick, Maine 04011}
\author{Panayotis G. Kevrekidis}
\affiliation{Department of Mathematics and Statistics, University of Massachusetts Amherst, Massachusetts 01003-4515}
\affiliation{Department of Physics, University of Massachusetts Amherst,  Massachusetts 01003}
\affiliation{Department of Mechanical Engineering, Seoul National University, 1 Gwanak-ro, Gwanak-gu, Seoul 08826, South Korea}
\author{Jinkyu Yang}
\email[]{jkyang11@snu.ac.kr}
\affiliation{Department of Mechanical Engineering, Seoul National University, 1 Gwanak-ro, Gwanak-gu, Seoul 08826, South Korea}

\date{\today}


\begin{abstract}
Rogue waves, characterized by their abrupt and extreme localization in space and time, have evolved from maritime folklore to subjects of intense study across diverse fields, from hydrodynamics~\cite{Walker2004, Fedele2016, Chabchoub2011} and nonlinear optics~\cite{Solli2007, Dudley2014} to plasmas and condensed matter physics~\cite{Charalampidis2018a, Shen2017}.
In mechanical systems, however, experimental realization remains elusive despite theoretical and numerical predictions~\cite{Han2014, Charalampidis2018, Kashyap2023}.
This gap stems from the stringent requirements for controllable nonlinearity, the high-fidelity initialization of the system, and the necessity to overcome inherent energy dissipation.
Here, we report the experimental formation of mechanical rogue waves in a precisely engineered one-dimensional metamaterial lattice with tailored nonlinearity and minimal dissipative losses. 
Using a precision electromagnetic release system, we prescribe initial strain profiles that trigger a transition from dispersive decay to extreme wave focusing.
Our parametric analysis reveals that the emergence of these extreme events is strictly contingent upon a synergy between high nonlinearity and a broad spatial energy reservoir within the initial seed. Crucially, neither factor alone is sufficient to overcome dispersion and trigger the observed focusing.
These findings establish a robust platform for studying transient nonlinear wave focusing phenomena in mechanical systems and offer insights for harnessing extreme wave localization for applications such as energy harvesting, waveguiding, and mechanical signal processing.
\end{abstract}

\maketitle

\section*{Introduction}

A spontaneous focusing of energy into highly localized features, known as rogue or extreme events, represents one of the most compelling, yet elusive, events in wave dynamics.
Observed in contexts ranging from unpredictable oceanic surges~\cite{Walker2004, Fedele2016, Chabchoub2011, Haver2004, Adcock2011, Chabchoub2012a, Adcock2018, Chabchoub2020}, ultra-short optical pulses~\cite{Solli2007, Dudley2014, Kibler2016, Tikan2017, Baronio2018}, to plasmas and condensed matter physics~\cite{Ruderman2010, Sabry2012, Bains2014, Tolba2015, Charalampidis2018a, Shen2017, Hohmann2010}, these phenomena are profound manifestations of nonlinear dynamics.
Unlike energy localization engineered through structural heterogeneities such as defects~\cite{Yablonovitch1987, Khelif2003}, disorder~\cite{Anderson1958, Hu2008}, or topological boundaries~\cite{Kane2013, Susstrunk2015}, rogue events emerge from a fundamentally different mechanism: strong nonlinearity overriding dispersion in homogeneous media to trigger self-focusing instability~\cite{BertolaTovbis2013}.
Furthermore, while solitons~\cite{Yasuda2019, Ye2023}, kinks~\cite{Veenstra2024}, and breathers~\cite{Cuevas2009, Chong2014} rely on a persistent balance between nonlinearity and dispersion for steady propagation, rogue events are intrinsically transient and characterized by extreme localization in \textit{both} space and time.

Theoretical frameworks for these extreme events in discrete media have been established across a wide range of mathematical models, from variants of the discrete nonlinear Schr\"odinger equation~\cite{Maluckov2013, Hoffmann2018, Sullivan2020, Yan2012} and Hirota lattices~\cite{Ankiewicz2010, Wen2018} to quantum disordered~\cite{Buarque2023} and dissipative~\cite{priya} systems. 
Within the specific context of phononic lattices exemplified by the Fermi-Pasta-Ulam-Tsingou (FPUT) chain, numerical studies predict that a localized ``seed'' of energy can precipitate a gradient catastrophe, culminating in rogue wave formation~\cite{Charalampidis2018}.
This behavior stands in sharp contrast to its linear counterparts, where initial pulses undergo rapid, monotonic dispersion~(Fig.~\ref{fig:main_schematic}a).

Despite these predictions, observing such transient focusing in a controlled laboratory setting remains a formidable challenge. 
The primary obstacles stem from the difficulty of tailoring lattices with the requisite nonlinear potentials, alongside an inherent susceptibility to dissipative losses that often attenuates the signal before focusing can occur.
Even in ideal nonlinear and nondissipative systems, replicating the spatially extended initial conditions assumed in mathematical models is non-trivial.
Conventional boundary excitations, which are effective in generating traveling nonlinear waves like solitons~\cite{Deng2019b, Zhang2023}, typically fail to produce the clean initial seed required for coherent nonlinear evolution.
Instead, they often lead to premature thermalization or chaotic wave dynamics before the focusing instability can fully develop. 

Here, we address these bottlenecks by architecting a mechanical metamaterial platform, which offers precise control over both inter-particle nonlinearity and full-field strain profiles.
Integrated with a specialized electromagnetic release system, this approach enables the high-fidelity realization of sculpted initial conditions across the entire lattice domain in the form of a Gaussian wave packet. 
Accordingly, this experimental capability allows us to reproduce a discrete analogue of the semiclassical dispersive wave regime, where the emergence of mechanical rogue waves is triggered by a so-called gradient catastrophe~\cite{BertolaTovbis2013, Charalampidis2018}.
By tracking the spatiotemporal evolution of the lattice, we experimentally observe these extreme localization events prevailing over inherent dissipation.
Our results demonstrate that the transition from dispersive decay to spatiotemporal localization is uniquely governed by the amplitude and width of the initial Gaussian profile.
By mapping this two-parameter space through systematic experiments, we identify the critical regime for the emergence of rogue-focused states.
These observations are corroborated by discrete numerical simulations of both discrete FPUT-type models and the corresponding continuum limit described by the nonlinear Schr\"odinger equation (NLSE).
Ultimately, these findings demonstrate how tailored nonlinear dynamics can overcome the inherent dissipation of solid media, providing physical insights and pathways for the precise control and extreme focusing of mechanical energy.

\section*{System design and theoretical framework}
\begin{figure*}[htbp]
    \centering
    \includegraphics[width=183truemm]{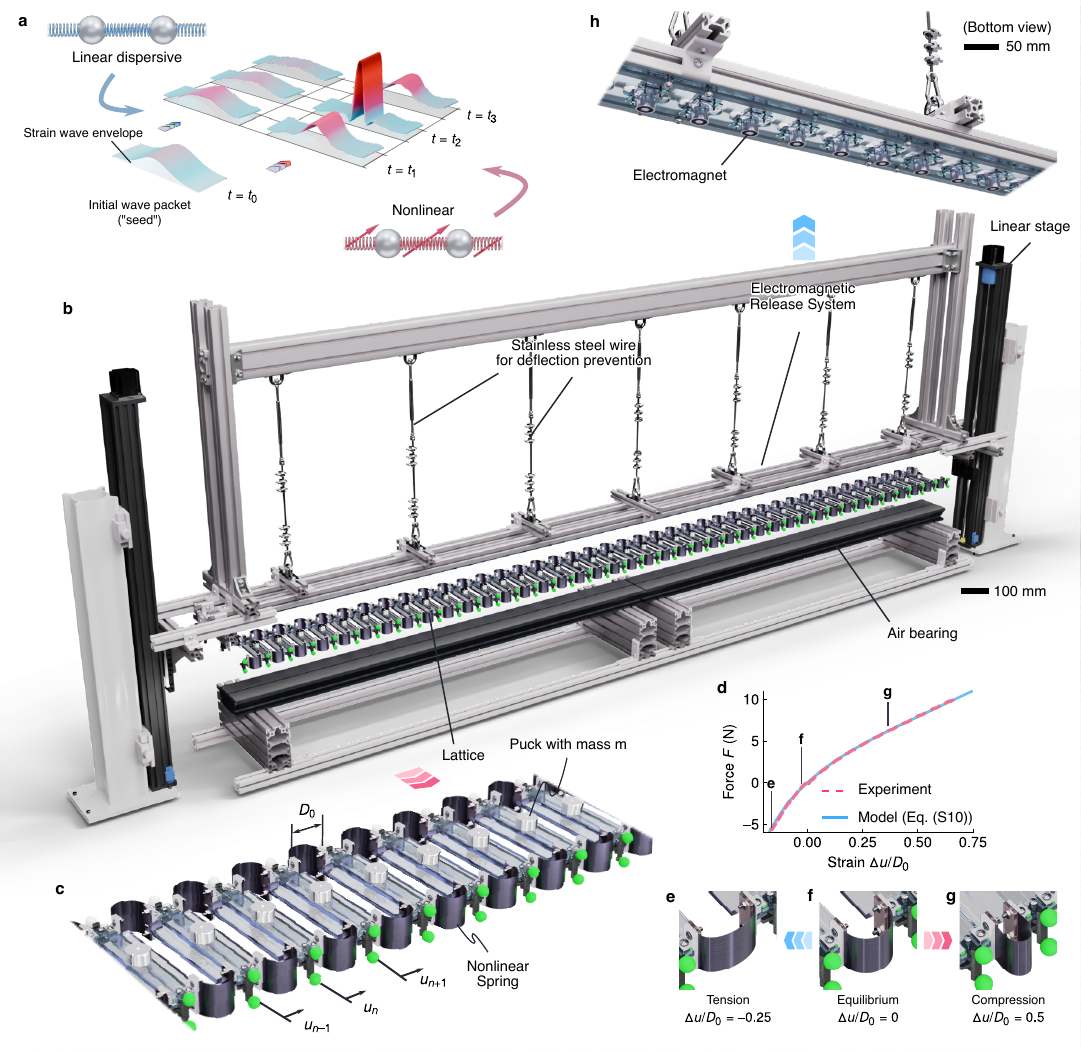}
    \caption{
    \textbf{Schematic illustration of wave focusing and experimental setup.}
    (\textbf{a}) Comparison of wave evolution in linear and nonlinear mechanical lattices, exhibiting dispersive decay and extreme focusing, respectively, from an identical initial perturbation (``seed'').
    (\textbf{b}) Experimental setup consisting of an electromagnetic release system, a nonlinear mechanical lattice, and an air bearing support.
    The release system is operated with linear stages, which can adjust the clearance between the lattice and the release mechanism with a precision of 5 $\mu$m resolution.
    (\textbf{c}) Detailed view of the lattice atop the air bearing. 
    The flat carrier (``puck'') behaves as a discrete mass, and a sheet of spring steel bent in semi-circular profile acts as a nonlinear spring.
    (\textbf{d}) The force-strain relationship of the nonlinear spring.
    Model (blue solid line; Eq.~(S10)) and experiment (red dashed line) show agreement.
    The deformed shape of the spring steel at
    (\textbf{e}) $\Delta u/D_0=-0.25$,
    (\textbf{f}) $0.0$, and
    (\textbf{g}) $0.5$.
    (\textbf{h}) The electromagnetic release system features an array of independently adjustable electromagnets that can be longitudinally displaced to encode a prescribed initial strain profile.
    }
    \label{fig:main_schematic}
\end{figure*}

Our experimental platform~(Fig.~\ref{fig:main_schematic}b, Extended Data Fig.~\ref{fig:ex_platform_overview},\ref{fig:ex_platform_component}) integrates three primary components: a nonlinear mechanical metamaterial lattice, an electromagnetic release system, and an air-bearing support structure~(Methods, Supplementary Note~1 and Supplementary Video~1).
Unlike typical lattice supporting structures, such as shafts or rollers, the air-bearing significantly reduces the friction between the lattice and the support structure, leaving only minimal intersite dissipation dominant~(Supplementary Note~2, Supplementary Video~2).
The lattice comprises 41 nonlinear spring units connecting ``pucks'' that travel on the air bearing~(Fig.~\ref{fig:main_schematic}c).
Each spring unit is made of a thin spring-steel sheet bent into a semi-circular profile.
Longitudinal deformation of these units yields the force-strain relationship shown in Fig.~\ref{fig:main_schematic}d, with corresponding states depicted in Fig.~\ref{fig:main_schematic}e--g.
Crucially, the unit cell exhibits a strain-softening response under compression~(Fig.~\ref{fig:main_schematic}g), and a strain-hardening response under tension~(Fig.~\ref{fig:main_schematic}e).
The reduced-order spring model, implemented in our analysis for explicit computation, captures this asymmetric behavior, showing excellent agreement with experimental uniaxial loading tests~(Supplementary Note~3).

The dynamics of our lattice can be approximated by the FPUT framework with cubic nonlinearity:
\begin{align}\label{eq:fput_eom}
    m_n\ddot{u}_n+\sum_{i=1}^{3}\alpha_{i+1}\left[(\Delta u_n)^i-(\Delta u_{n-1})^i\right]=0,
\end{align}
with $\Delta u_n=u_n-u_{n+1}$ denoting the deformation of the $n$-th spring unit.
A multiple-scale analysis~(Supplementary Note~4) reduces this discrete system to the NLSE:
\begin{align}\label{eq:nlse}
    i\partial_\tau{A_{1,1}}+\nu_2\partial_\xi^2{A_{1,1}}+\nu_3{A_{1,1}}|{A_{1,1}}|^2=0,
\end{align}
which describes the evolution of the wave packet envelope, $A_{1,1}$, in a Hamiltonian scenario at slow scale $\xi=\epsilon(n-\lambda t)$ and $\tau=\epsilon^2t$ with $\lambda$ being the group velocity.
In the context of Hamiltonian dynamics, the strain-softening nature of our unit cells under compression ensures that the resulting NLSE is of a focusing type~($\nu_2<0$, $\nu_3>0$), a condition that triggers a gradient catastrophe for a sufficiently wide initial wave packet~\cite{Charalampidis2018}.

The electromagnetic release system~(Fig.~\ref{fig:main_schematic}h, Extended Data Fig.~\ref{fig:ex_platform_component}, \ref{fig:ex_circuit}) encodes an initial wave packet.
Each mobile puck is paired with a stationary electromagnet; by longitudinally displacing these electromagnets to match a target spatial profile and subsequently activating them, the entire lattice is magnetically captured and held in a precisely pre-strained state.
This methodology allows ``dialing in'' prescribed initial conditions to the lattice.
In the present work we impose a Gaussian-type wave packet~\cite{Charalampidis2018} at the edge of the Brillouin zone~($k=\pi$):
\begin{align}\label{eq:gaussian}
    \Delta u_n|_{t=0} = P_0\exp\left[-\left(\dfrac{n}{2\sigma}\right)^2\right]\left(-1\right)^n.
\end{align}
In the above formulation, $P_0$ denotes the scaling factor dictating the initial nonlinearity, and $\sigma$ denotes the width of the wave packet governing the initial concentration of energy in space.
Unlike canonical analytical rogue wave solutions, such as the Peregrine soliton, the Gaussian-type seed naturally decays to zero at the boundaries.
This property ensures physical realizability within a finite domain and mitigates boundary-induced artifacts or global extension or contraction of the entire chain.

Encoding both narrow and wide Gaussian seeds with identical initial amplitude~($P_0$) yields \textit{in situ} strain profiles that exhibit very good agreement with the analytical envelope prescribed by Eq.~\eqref{eq:gaussian} (Fig~\ref{fig:main_strainfield}a,b).
This high-fidelity seed establishes a precisely tunable platform, allowing us to address a fundamental question: Does extreme energy focusing favor a narrow seed due to its pre-existing spatial localization, or a wider profile that harnesses a larger reservoir of potential energy across the lattice?

\section*{Experimental observation of spatiotemporal localization}
\begin{figure*}[htbp]
    \centering
    \includegraphics[width=183truemm]{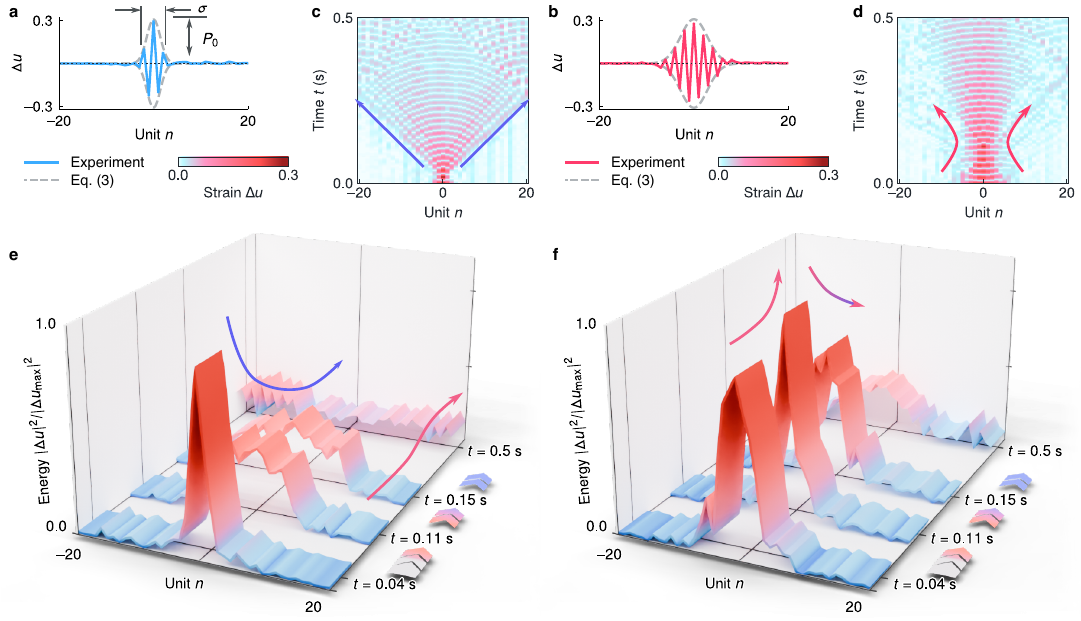}
    \caption{
    \textbf{Experimental strain landscape in focusing and non-focusing scenarios.}
    (\textbf{a}) Initial Gaussian profile of the non-focusing case ($P_0=0.31$, $\sigma=1.0$).
    (\textbf{b}) Initial Gaussian profile of the focusing case ($P_0=0.31$, $\sigma=2.0$).
    (\textbf{c}) 2D surface map of the non-focusing case showing a circular sector dispersive profile.
    (\textbf{d}) 2D surface map of the focusing case showing hourglass-like focusing profile.
    (\textbf{e}) Strain snapshot of the non-focusing case showing the rapid decrease of the initial packet amplitude and increase of the background wave.
    (\textbf{f}) Strain snapshot of the focusing case showing an increase in initial wave packet amplitude before a gradual decay.
    }
    \label{fig:main_strainfield}
\end{figure*}

We experimentally investigate the lattice dynamics initiated by the narrow~($\sigma=1.0$; Fig.~\ref{fig:main_strainfield}a) and wider~($\sigma=2.0$; Fig.~\ref{fig:main_strainfield}b) Gaussian seeds with identical initial amplitude ($P_0=0.31$).
Experimental procedures are detailed in the Methods and Supplementary Video~3.
In the first case, the narrow seed undergoes rapid decay, dispersing the concentrated energy across the lattice~(Fig.~\ref{fig:main_strainfield}c).
This rapid dilution indicates that for highly localized seeds, dispersion remains the dominant mechanism.
On the contrary, the wider seed exhibits pronounced self-focusing, consistent with the semiclassical regime observed in both continuum~\cite{BertolaTovbis2013} and discrete FPUT~\cite{Charalampidis2018} frameworks.
The wave packet concentrates toward the lattice center before eventually dispersing, producing a distinct hourglass-shaped landscape~(Fig.~\ref{fig:main_strainfield}d), a hallmark of spatiotemporal localization and of the gradient catastrophe scenarios~\cite{BertolaTovbis2013, Charalampidis2018}.
Notably, the energy remains evanescently confined near the center even after the primary focusing phase, effectively suppressing propagation toward the boundaries over the duration of the experiment and highlighting a regime where nonlinearity significantly delays dispersive spreading.

Extracting the instantaneous spatial strain profiles captures the stark difference of these two paths~(Fig.~\ref{fig:main_strainfield}e,f, Supplementary Video~4).
The narrow seed~($\sigma=1.0$) broadens monotonically, accompanied by a rapid decay of the envelope amplitude.
The transition to a plane wave-like profile by $t=0.5$~s confirms the dominance of dispersive effects, despite the initial nonlinearity~(Fig.~\ref{fig:main_strainfield}e).
In contrast, the wider initial packet ($\sigma=2.0$) exhibits evident self-amplification, reaching its peak amplitude when its spatial width is most compressed at $t=0.11$~s and thus, manifesting the spatiotemporal wave focusing~(Fig.~\ref{fig:main_strainfield}f). 
Notably, this localized state leaves a pronounced spatiotemporal localization signature; even at $t=0.5$~s, the strain amplitude at the central unit ($n=0$) remains significantly elevated compared to the boundaries (Supplementary Video 3).
This spatiotemporal localization clearly distinguishes itself from the purely dispersive lattice dynamics.

\begin{figure*}[htbp]
    \centering
    \includegraphics[width=183truemm]{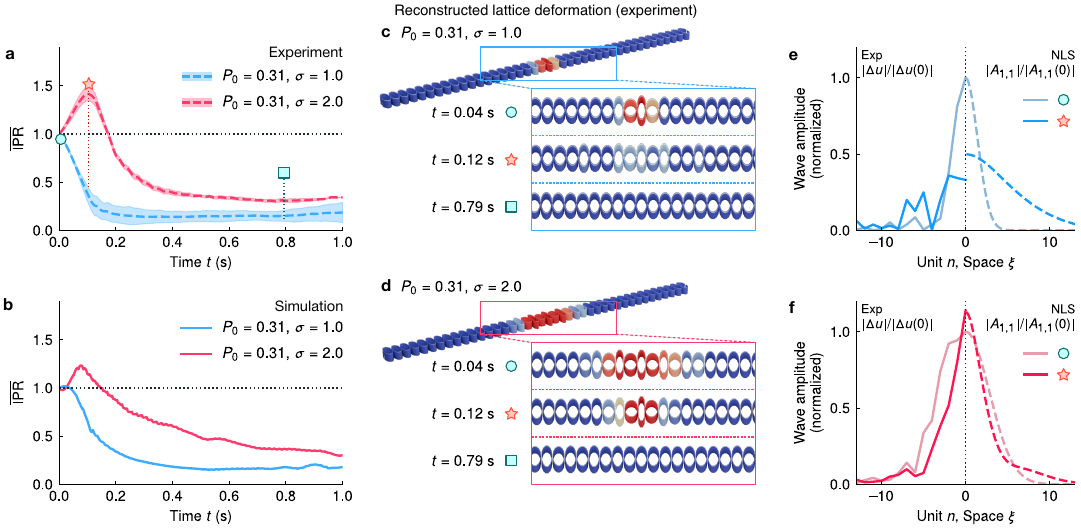}
    \caption{
    \textbf{Localization characterization.}
    Normalized inverse participation ratio ($\overline{\mathrm{IPR}}$) for narrow ($\sigma = 1.0$) and wide ($\sigma = 2.0$) case as a function of time in the
    (\textbf{a}) experiment and
    (\textbf{b}) simulation.
    Experimental data are the average of five trials~(dashed lines) with shaded regions denoting the standard deviation.
    (\textbf{c}, \textbf{d}) Reconstructed lattice deformation at three sequential stages.
    (\textbf{c}) In the non-focusing case ($\sigma=1.0$), the wave packet spreads monotonically over time.
    (\textbf{d}) In the focusing case ($\sigma=2.0$), the center units exhibit a contraction, concentrating energy into a highly localized cluster before dispersing.
    (\textbf{e},\textbf{f}) Wave amplitude normalized by the initial peaks for experiment (left-half; solid lines) and NLS simulation (right-half; dashed lines).
    The pale and thick colors correspond to the times $t=0.04$ and $0.12$ s (denoted with circle and star symbols in the other panels), respectively.
    (\textbf{e}) In the non-focusing case ($\sigma=1.0$).
    (\textbf{f}) In the focusing case ($\sigma=2.0$).
    }
    \label{fig:main_iprtemporal}
\end{figure*}

To quantify the temporal evolution of wave localization, we monitor the inverse participation ratio (IPR), a metric for localization in lattice dynamics~\cite{PRB_IPR}.
For a direct comparison across varying initial widths, we define a normalized IPR relative to the initial state:
\begin{align}\label{eq:ipr}
    \overline{\mathrm{IPR}}(t)=
    \frac{\mathrm{IPR}(t)}{\mathrm{IPR}(0)},
    \qquad
    \mathrm{IPR}(t)=\frac{\sum_{i=1}^{N}\left|\Delta u_i(t)\right|^4}{\left[\sum_{j=1}^{N}\Delta u_j^2(t)\right]^2}.
\end{align}
A higher $\overline{\mathrm{IPR}}$ indicates a more spatially localized strain distribution. 
Tracking $\overline{\mathrm{IPR}}$ over time reveals a distinct temporal signature between the non-focusing and focusing cases~(Fig.~\ref{fig:main_iprtemporal}a; blue and red dashed lines, respectively).
In the non-focusing case (blue), the monotonic decrease in $\overline{\mathrm{IPR}}$ aligns with the observed dispersive broadening; specifically, the rapid drop within the first $0.1$~s quantitatively captures the simultaneous amplitude attenuation and width expansion.
In contrast, the focusing case (red) exhibits an abrupt increase in $\overline{\mathrm{IPR}}$ that surpasses the initial unity value.
This surge peaks near $t=0.1$~s, corresponding to the formation of the maximally compressed wave packet with the highest localized amplitude.
Numerical simulations incorporating the reduced-order spring model corroborate these experimental findings~(Fig.~\ref{fig:main_iprtemporal}b).

These focusing and dispersive dynamics described by $\overline{\mathrm{IPR}}$ are faithfully reflected in the reconstructed lattice deformations~(Fig.~\ref{fig:main_iprtemporal}c,d).
While the dispersive packet attenuates rapidly~(Fig.~\ref{fig:main_iprtemporal}c), the focusing case undergoes a pronounced physical contraction, funneling the energy into a localized cluster of approximately 3--5 units before eventual dispersion~(Fig.~\ref{fig:main_iprtemporal}d).
Direct comparison of the spatial waveform at peak focusing with numerical NLS predictions confirms that our system captures a robust extreme spatiotemporal localization of rogue waves~(Fig.~\ref{fig:main_iprtemporal}e,f).
Beyond spatial focusing, the spatial spectral analysis resolves the canonical triangular exponential decay~\cite{Akhmediev2011a, Kibler2010, Chabchoub2012} (Supplementary Note~5), confirming the rogue wave nature of the experimental measurement.
This robust localization emerges despite inherent dissipation that typically suppresses the physical observation of theoretically predicted amplification~\cite{amin}.
Crucially, these comparisons suggest that high nonlinearity alone is insufficient to trigger localization; the initial seed must also possess sufficient spatial breadth to provide an energy reservoir.
In this semiclassical regime, this reservoir facilitates the gradient catastrophe~\cite{BertolaTovbis2013, Charalampidis2018} required to precipitate the observed rogue wave phenomenon.

\section*{Initial condition parameter space and localization}

\begin{figure*}[htbp]
    \centering
    \includegraphics[width=183truemm]{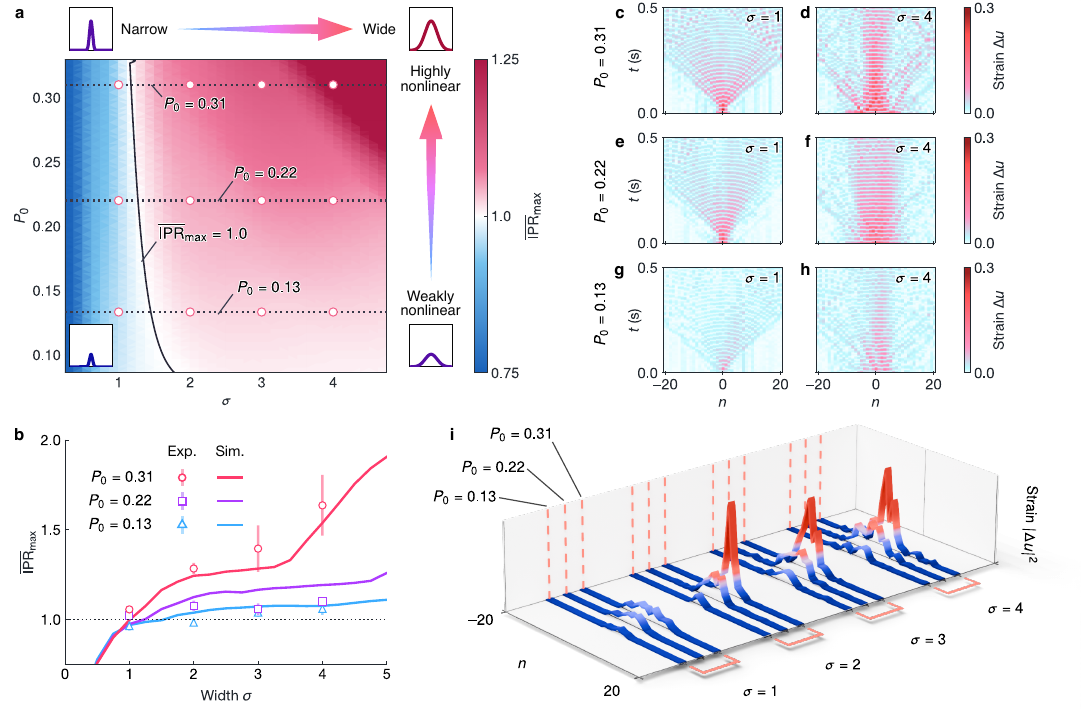}
    \caption{
    \textbf{Parametric study of nonlinear wave focusing.}
    (\textbf{a}) Variation of $\overline{\mathrm{IPR}}_\mathrm{max}$ in an initial condition parameter space $(P_0,\,\sigma)$.
    (\textbf{b}) $\overline{\mathrm{IPR}}_\mathrm{max}$ as a function of $\sigma$ for three $P_0$ values: $0.31$ (red), $0.22$ (purple), and $0.13$ (blue).
    Open symbols, experiment; solid lines, simulation.
    Strain wave field of the representative cases from the experiment with
    $P_0=0.31$,
    (\textbf{c}) $\sigma=1.0$,
    (\textbf{d}) $\sigma=4.0$;
    $P_0=0.22$,
    (\textbf{e}) $\sigma=1.0$,
    (\textbf{f}) $\sigma=4.0$;
    $P_0=0.13$,
    (\textbf{g}) $\sigma=1.0$,
    (\textbf{h}) $\sigma=4.0$.
    (\textbf{i}) Spatial strain profiles from the experiment under various combinations of $P_0$ and $\sigma$.
    The focusing cases are taken at the moment of highest wave amplitude, while for the non-focusing case, the profiles are extracted at the same $t$ as the corresponding focusing case.
    }
    \label{fig:main_localization}
\end{figure*}

To systematically probe the interplay of the parameters, we map the maximum $\overline{\mathrm{IPR}}$, which we call $\overline{\mathrm{IPR}}_{\mathrm{max}}$, over the duration of the experiment $t\in[0.0,1.0]$ s, across the parameter space of initial amplitude ($P_0$) and width ($\sigma$) via experiment and lattice numerical simulations~(Fig.~\ref{fig:main_localization}a). 
In this parameter space, $\overline{\mathrm{IPR}}_{\mathrm{max}}$ remains low for narrow seeds across all amplitudes considered. 
Even with elevated $P_0$, the region near $\sigma \approx 1$ fails to focus ($\overline{\mathrm{IPR}}_{\mathrm{max}} < 1$), whereas wider seeds ($\sigma \geq 2$) readily transition into the focusing regime ($\overline{\mathrm{IPR}}_{\mathrm{max}} > 1$). 
This is consistent with our experimental observations.
Notably, the significantly higher $\overline{\mathrm{IPR}}_{\mathrm{max}}$ values in the upper-right quadrant visually confirm the synergistic requirement: triggering wave focusing demands both high initial amplitude and sufficiently broad spatial reservoir.

We augment this numerical landscape with 12 targeted experimental measurements, encompassing all combinations of $P_0\in\{0.13,0.22,0.31\}$ and $\sigma\in\{1.0,2.0,3.0,4.0\}$~(Fig.~\ref{fig:main_localization}a, open circles).
While most cases belong to the transition region~($\overline{\mathrm{IPR}}_{\mathrm{max}}\approx1$), the most extreme case ($P_0=0.31, \sigma=4.0$) resides within the high-$\overline{\mathrm{IPR}}$ focusing regime.
Numerical cross-sections extracted at respective $P_0$ reveal agreement between the experimental data and the simulation results~(Fig.~\ref{fig:main_localization}b).
Consistent with the underlying model, increasing $P_0$ generally elevates $\overline{\mathrm{IPR}}_{\mathrm{max}}$ for a given $\sigma$, an effect that remains discernible even in the dispersion-prevailing $\sigma=1.0$ regime.
The most striking difference, however, emerges at $P_0=0.31$.
While the lower amplitude cases ($P_0=0.13, 0.22$) exhibit a plateaued response regardless of width, the $P_0=0.31$ case undergoes a sharp nonlinear escalation beyond $\sigma=3.0$, marking the clear threshold where nonlinearity dominates dispersive broadening, despite the presence of the dissipation.

The spatiotemporal strain evolution of the representative cases directly visualizes these distinct dynamics~(Fig.~\ref{fig:main_localization}c--h).
Across all tested amplitudes, the narrow-seed cases ($\sigma=1.0$) consistently exhibit rapid dispersive decay, characteristic of the low-$\overline{\mathrm{IPR}}_{\mathrm{max}}$ regime (Fig.~\ref{fig:main_localization}a, blue region).
In contrast, wider packets ($\sigma=4.0$) bifurcate in their dynamics depending on the nonlinearity.
At lower amplitudes~($P_0=0.13$ and $0.22$), the initial packet either undergoes dispersive decay or maintains its width without focusing through $t=0.5$~s.
In the high-nonlinearity case~($P_0=0.31$), however, sufficiently wide 
initial wavepacket exhibits rapid contraction toward $t \approx 0.15$~s and remains localized until $t=0.5$~s, signaling the reproducible extreme spatiotemporal localization of the rogue waves.

Evaluating the spatial energy density ($\left| \Delta u \right|^2$) at the instant of peak localization quantifies this energy distribution across the sampled parameter space (Fig.~\ref{fig:main_localization}i).
To ensure a consistent comparison, snapshots for non-focusing regimes are extracted at the exact time-steps as their focusing counterparts of the same initial amplitude $P_0$.
At the lowest amplitude~($P_0=0.13$), the profiles exhibit only minor perturbations near the lattice center, a clear signature of dispersive behavior.
While moderate amplitudes ($P_0=0.22$) induce a slight energy concentration, the energy distribution remains inherently spread, failing to achieve evident focusing.
At $P_0=0.31$, however, the strain energy transitions beyond mere spatial confinement, culminating in an acute, dominant central peak.
This distinct divergence between low- and high-$P_0$ regimes reconfirms the necessity of strong nonlinearity for the emergence of mechanical rogue waves.

\section*{Conclusions \& Outlook}
In summary, we have experimentally demonstrated the formation of mechanical rogue waves in a nonlinear lattice, originating from the energy focusing scenario that governs the extreme spatiotemporal localization of semiclassical wave packets.
Our integrated experimental platform, featuring a one-dimensional lattice with tailored strain-softening nonlinearity, an electromagnetic initialization system, and an air bearing substrate, effectively mitigates inherent dissipation while enabling high-fidelity control over the prescribed Gaussian wave packet.
Through a systematic parametric study of a full two-parameter plane of initial wavepacket amplitudes and widths, we have revealed that the high nonlinearity alone is insufficient for the formation of such extreme events; rather, it is the interplay between the initial amplitude and width of the packet that dictates the transition from dispersive decay to extreme wave focusing.
While narrow seeds tend to be dominated by dispersion regardless of amplitude, wider seeds provide the necessary spatial energy reservoir that overcomes dispersion when coupled with sufficient nonlinearity.
These findings establish a versatile framework for predictable generation and control of extreme localization in mechanical settings, extending the study of extreme nonlinear wave phenomena, traditionally observed in fluid or optical counterparts~\cite{Chabchoub_water1, jshe}, into the domain of architected materials.
Furthermore, our study paves the way for a novel class of engineered materials designed to harness the targeted localization of high-density energy, showing potential in applications for resilient energy systems and mechanical information technology.

\clearpage

\section*{Methods}

\subsection*{Fabrication and lattice assembly}
The experimental platform comprises a one-dimensional nonlinear lattice, an electromagnetic release array, and a low-friction air-bearing substrate (New Way Air Bearings, Transition Air Bar, $1250\times100$ mm connected in series), supported by a pneumatic regulation subsystem and a synchronized dual high-speed camera system~(Fig.~\ref{fig:main_schematic}, Extended Data Fig.~\ref{fig:ex_platform_overview}).
The air bearing is elevated on aluminum frames to provide sufficient clearance from the optical table, accommodating the necessary pneumatic routing operating at 145 kPa. 
To maintain the pristine condition of the porous media within the air bearing, the pneumatic line incorporates in-line filtration and desiccant drying stages, precluding particulate clogging and internal oxidation.

The lattice consists of 41 unit cells, each featuring a semi-circular spring steel component ($0.1$-mm-thick SK5) cut using a laser cutter (Bodor i6 Fiber Laser) to dimensions of $30\times30\pi$~mm.
The bending of these units provides the strain-softening nonlinearity under compression required for energy focusing and serves as the mechanical coupling between the particle carriers, or pucks~(Extended Data Fig.~\ref{fig:ex_platform_component}a,b).
Each carrier is constructed from two 5-mm-thick acrylic plates (carrier base and spacer) processed with a laser cutter (Totec Speedy 360) stacked together, which support a 3D-printed integrated L-bracket and a threaded carbon steel EM pole counterpart for magnetic capture.
For full-field motion tracking, 10-mm diameter polytetrafluoroethylene (PTFE) spheres are coated with fluorescent green lacquer and mounted to each carrier via 3D-printed marker pins.
Once assembled, the 41-unit chain is placed atop the pressurized air bearing to minimize onsite damping.

The electromagnetic release system, consisting of 40 independent electromagnets aligned with the metallic block of the carriers (Fig.~\ref{fig:main_schematic}h, Extended Data Fig.~\ref{fig:ex_platform_component}c,d), initializes the lattice.
The electromagnets are fixed onto acrylic array holders, which are then mounted on an aluminum extruded frame. 
The acrylic holder and stabilizing washers fix the electromagnet tightly onto the array holder, preventing the electromagnet from tilting when a horizontal force is exerted by the lattice. 
Each electromagnet is allowed to move along the slit of the acrylic holder with reference scales to position the electromagnet in place.
To ensure structural homogeneity and maintain a uniform clearance between the electromagnet poles and the lattice across the 2-m track, the releaser frame is stabilized against gravitational deflection using an optimized 7-point stainless steel wire suspension system.
For a comprehensive description of the system geometry, tolerance, and assembly protocols, see Supplementary Note~1 and Supplementary Videos~5 and 6.

\subsection*{Experimental procedures}

The initial displacement field $u_n$ is established by prescribing the longitudinal displacements of the 40 mobile particles. 
To generate the target Gaussian precursor, the displacement $u_n$ is calculated as the discrete cumulative sum of the target strain field defined in Eq.~\eqref{eq:gaussian}, accounting for the zone-boundary mode ($k=\pi$):
\begin{equation}\label{eq:initia_disp}
    u_n|_{t=0} = \sum_{j=n}^{20} P_0 \exp \left[ -\left( \frac{j}{2\sigma} \right)^2 \right] \cos(kj) - \mathcal{C}.
\end{equation}
where $P_0$ is the peak strain amplitude, and $\mathcal{C}$ is a centering constant that ensures a zero-mean displacement across the lattice ($\sum u_n = 0$).
Based on Eq.~\eqref{eq:initia_disp}, each electromagnet on the releaser array is manually programmed to the calculated coordinates $u_n$.
To preserve the fidelity of encoded Gaussian profiles prior to release, the holding force of each electromagnet is individually tuned to counterbalance the local nonlinear restoring forces of the strained lattice.
The 40 electromagnets are segmented into three independent power domains (to prevent voltage drop) and integrating a regulating resistor in series with each particle (Extended Data Fig.~\ref{fig:ex_circuit}a,b).
The resulting spatial distribution of the manually calibrated electromagnetic power mirrors the required holding force to maintain the initial wave packet (Extended Data Fig.~\ref{fig:ex_circuit}c--e; Supplementary Note~6, Supplementary Tables~1,2 for detailed circuit parameters).
This force-balancing protocol ensures high-fidelity and reproducible strain profiles across all tested values of $P_0$ and $\sigma$ (Extended Data Fig.~\ref{fig:ex_initial_state}).
The entire release assembly is subsequently lowered until a vertical clearance of approximately $100~\mu$m~(measured using a shim sheet) is achieved between the electromagnet poles and the metallic blocks of the particles.
The electromagnet relay circuit is then activated to capture the particles.
To ensure that the initial potential energy is stored exclusively within the prescribed longitudinal strain field, each particle is manually centered relative to its corresponding electromagnet pole.
This step is critical to eliminate lateral displacement errors and misalignments.
Once initialized, the air-bearing system is pressurized via a high-pressure compressor.

The wave evolution is initiated by a simultaneous power cutoff to the electromagnets via the relay circuit, ensuring synchronized triggers across all 40 particles with minimal temporal jitter.
The resulting spatiotemporal evolution is captured by a pair of synchronized high-speed cameras~(Kron Technologies Chronos 1.4) tracking fluorescent green markers at $2{,}000$ frames-per-second.
The acquisition rate is orders of magnitude higher than the carrier wave frequency at the edge of the Brillouin zone, providing sufficient temporal resolution to resolve the gradient catastrophe or wave focusing.
For more details of experiment procedures, see Supplementary Note~1 and Supplementary Video 2.

\subsection*{Numerical methods}

Lattice dynamics are simulated by integrating the equations of motion using an eighth-order Dormand-Prince method with a maximum time step of $10^{-6}$~s to ensure numerical stability and energy conservation.
To complement these discrete results, the Nonlinear Schr\"odinger equations are solved using a sixth-order central difference scheme for spatial discretization, combined with the same eighth-order Dormand-Prince integrator for time-stepping. 
Both integration schemes incorporate adaptive error estimation of the fifth and third orders.
Additionally, the static configuration for the elastic model is determined via a shooting method utilizing an initial value problem solver. 
All computational procedures are implemented in double-precision floating-point format, with specific simulation parameters provided in Supplementary Table~3.

\section*{Data availability}
Data supporting the findings of this study are available in the Zenodo repository under the accession code: {https://doi.org/10.5281/zenodo.18884773}.

\section*{Acknowledgments}
We thank Dr. Amin Chabchoub (Okinawa Institute of Science and Technology) for helpful discussion. Y.M. and J.Y. acknowledge the support from SNU-IAMD, SNU-IOER, US National Science Foundation [CMMI-1933729], and National Research Foundation grants funded by the Korean government [2023R1A2C2003705 and 2022H1D3A2A03096579].

\section*{Author contributions}
Y.M., C.C., P.G.K., and J.Y. conceptualized the work;
Y.M. performed analytical calculations, numerical simulations, and experiments, and drafted the manuscript;
C.C. and P.G.K. provided the analytical and computational guidance; J.Y. supervised the overall project.
All authors extensively contributed to the work and finalizing the manuscript.

\section*{Conflict of interest}
The authors declare no competing financial interest.


\bibliography{main}

\clearpage
\renewcommand{\figurename}{Fig.}

\clearpage
\setcounter{section}{0}
\setcounter{equation}{0}
\setcounter{figure}{0}
\setcounter{table}{0}
\renewcommand{\figurename}{Extended Data Fig.}

\begin{figure*}[htbp]
    \centering
    \includegraphics[width=183truemm]{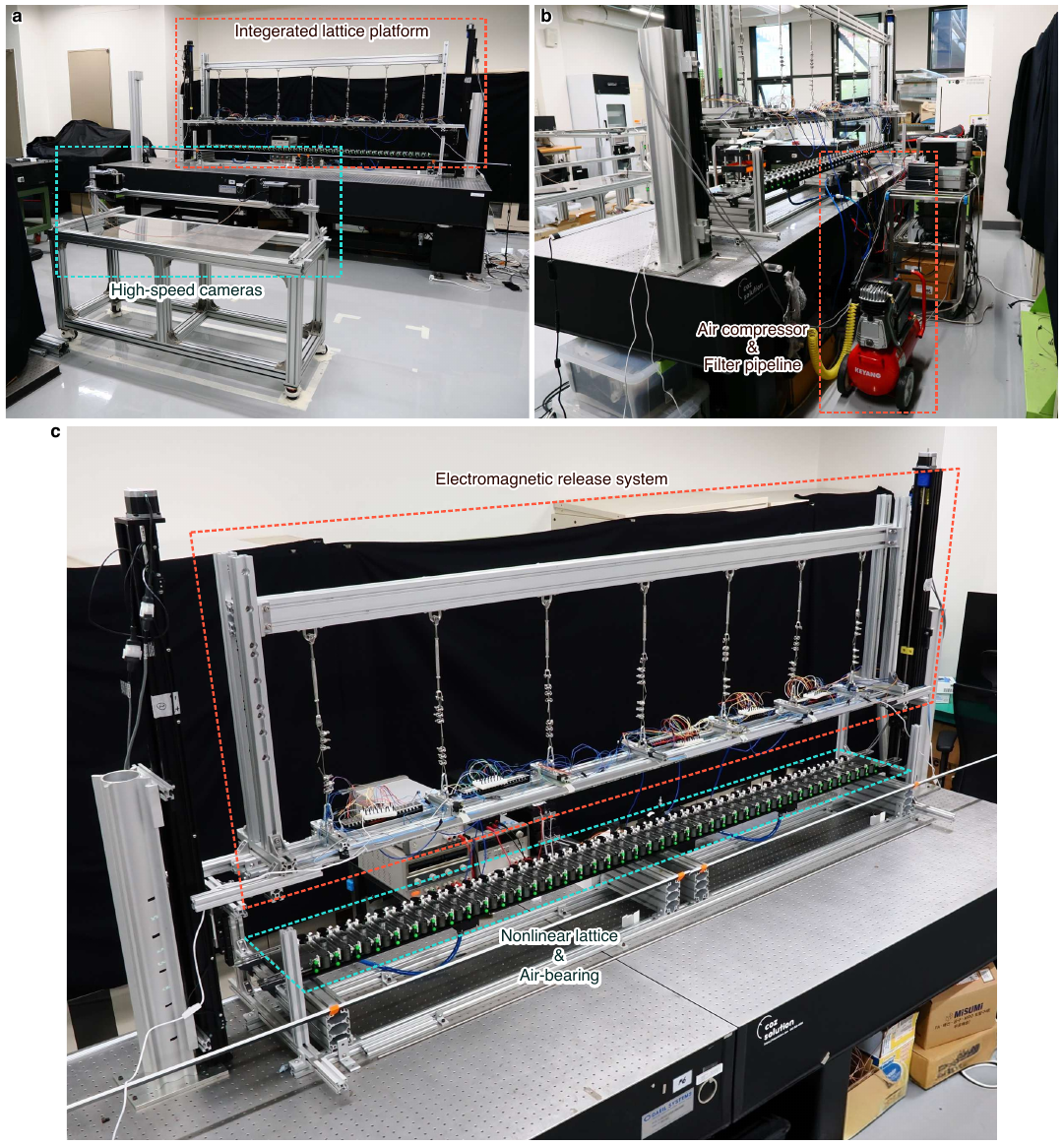}
    \caption{
    \textbf{Experimental platform for the observation of mechanical rogue waves.}
    (\textbf{a}) Global view of the optical bench and camera mount configuration. 
    A synchronized dual high-speed camera system is positioned in front of the bench to capture the full-field spatiotemporal evolution of the lattice markers.
    (\textbf{b}) Pneumatic subsystem located at the rear of the platform. An air compressor and high-precision filtration pipeline ensure a steady supply of regulated air to the air-bearing substrate, minimizing non-conservative damping. 
    (\textbf{c}) Close-up view of the integrated assembly, illustrating the spatial alignment between the electromagnetic release system and the one-dimensional nonlinear lattice.
    }
    \label{fig:ex_platform_overview}
\end{figure*}

\begin{figure*}[htbp]
    \centering
    \includegraphics[width=183truemm]{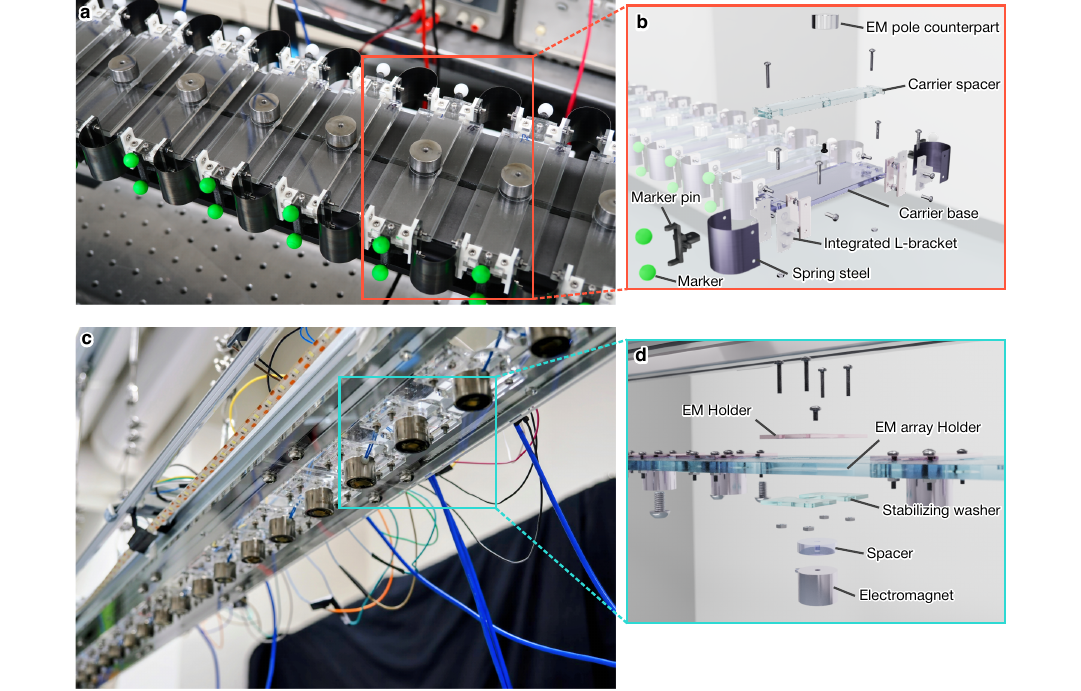}
    \caption{
    \textbf{Architecture of the metamaterial lattice and electromagnetic release array.}
    (\textbf{a}) Physical implementation and 
    (\textbf{b}) 3D-rendered exploded view of the nonlinear lattice. 
    Each unit consists of a discrete mass carrier (puck) connected via a semi-circular spring steel unit fixed onto integrated L-brackets, providing the tailored strain-softening nonlinearity required for energy focusing.
    The electromagnetic (EM) pole counterpart, with a polished surface, is magnetically held by an electromagnet.
    Markers are mounted on pins for motion tracking.
    (\textbf{c}) Photographic and 
    (\textbf{d}) schematic exploded-view representations of the electromagnetic release system. 
    The assembly utilizes an EM array holder and precision spacers to ensure a uniform 100 $\mu$m clearance between the electromagnet poles and the lattice pucks. 
    }
    \label{fig:ex_platform_component}
\end{figure*}

\begin{figure*}[t]
    \centering
    \includegraphics[width=183truemm]{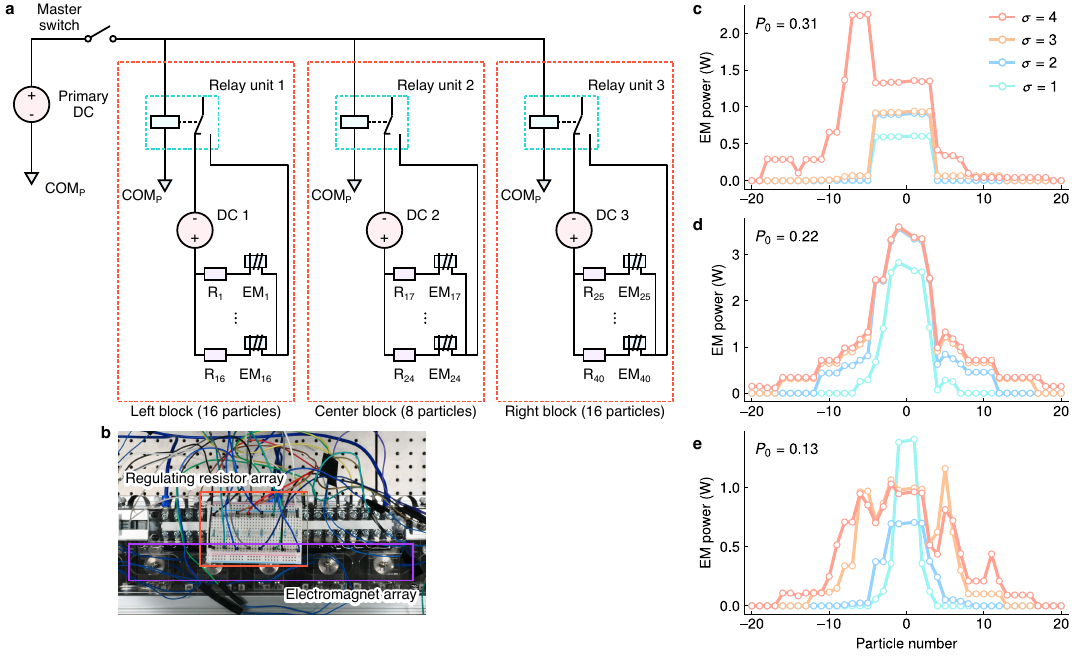}
    \caption{\textbf{Architecture and power distribution of the electromagnetic release system.}
    (\textbf{a})Circuit schematic illustrating the segmented three-block architecture (Left: 16 units; Center: 8 units; Right: 16 units). A primary DC supply and master toggle switch provide a synchronized trigger signal to three independent relay units, ensuring the simultaneous release of all 40 particles. Each block is powered by a dedicated DC supply to maintain voltage stability across the array during the capture phase. Individual regulating resistors ($R_n$) are connected in series with the corresponding electromagnets ($\mathrm{EM}_n$) to modulate the local magnetic holding force.
    (\textbf{b}) Photographic representation of the experimental assembly, highlighting the regulating resistor array on the breadboard and the underlying electromagnet array positioned atop the air-bearing support.
    (\textbf{c}--\textbf{e}) Spatial distribution of the calculated power supplied to each electromagnet for initial amplitudes of \textbf{c}, $P_0 = 0.31$, \textbf{d}, $P_0 = 0.22$, and \textbf{e}, $P_0 = 0.13$. The power values are derived from the individual regulating resistances and measured coil resistances ($R_\mathrm{EM} = 47.51 \pm 0.78\ \Omega$). The distribution is tailored to counterbalance the spatially varying mechanical restoring forces of the Gaussian initial conditions, with power levels peaking at the lattice center to host the extreme wave focusing events. The different curves in each panel correspond to initial widths of $\sigma = 1$ (cyan), $2$ (blue), $3$ (orange), and $4$ (red). 
    }
    \label{fig:ex_circuit}
\end{figure*}

\begin{figure*}[htbp]
    \centering
    \includegraphics[width=183truemm]{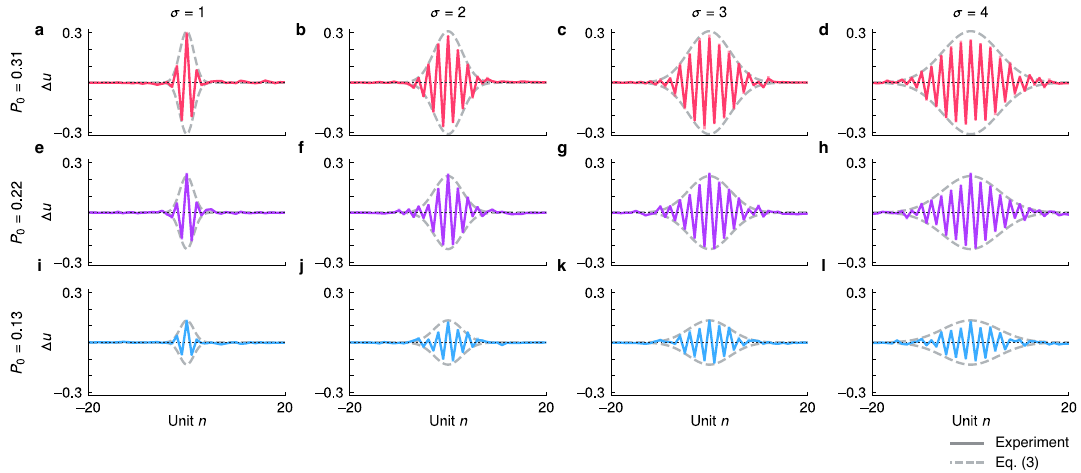}
    \caption{
    \textbf{Validation of initial state fidelity across the parameter space.}
     (\textbf{a}--{l}) Comparison of the initial experimental strain $\Delta u$ (solid colored lines) against the target analytical profiles (gray dashed lines) prescribed by Eq.~\eqref{eq:gaussian}.
     The grid represents the combination of peak amplitudes $P_0 \in \{0.31, 0.22, 0.13\}$ and spatial widths $\sigma \in \{1.0, 2.0, 3.0, 4.0\}$ utilized for the parametric study.
     The high degree of spatial overlap across all 12 cases demonstrates the precision of the electromagnetic release system in mapping mathematically prescribed wave packets onto the lattice.
     These high-fidelity ``seeds'' ensure that the subsequent wave evolution is driven by the intrinsic nonlinear dynamics of the system rather than initialization artifacts.
    }
    \label{fig:ex_initial_state}
\end{figure*}


\end{document}